\documentclass[epj,twocolumn]{svjour}
\usepackage{graphicx,multicol,color}
\begin{document}
\title{Modulational instability, rogue waves, and envelope solitons in opposite polarity dusty plasmas}
\author{M. H. Rahman$^*$, N. A. Chowdhury, A. Mannan,  M. Rahman, \and   A. A. Mamun}
\institute{Department of Physics, Jahangirnagar University, Savar, Dhaka-1342, Bangladesh\\
E-mail:   rahman1992phy@gmail.com$^*$}
\date{Received: date / Revised version: date}
\abstract{
  Dust-acoustic (DA)  waves (DAWs) and their modulational instability (MI)  have been investigated
  theoretically in a plasma system consisting of inertial opposite polarity (positively and negatively)
  warm adiabatic charged dust particles as well as inertialess non-extensive $q$-distributed electrons
  and non-thermal ions. A nonlinear Schr\"{o}dinger (NLS) equation is derived by using the reductive
  perturbation method. It has been observed from the analysis of NLS equaion that the  modulationally stable
  solitary DAWs give rise to the existence of dark envelope solitons, and that the modulationally unstable
  solitary DAWs give rise to the existence of bright envelope solitons or rogue structures. It is also
  observed for the fast mode of DAWs that the basic features (viz. stability of the DAWs,
  MI growth rate, amplitude and width of the DA rogue waves, etc.) are significantly modified by
  the related plasma parameters (viz. dust masses, dust charge state, non-extensive parameter $q$,
  and non-thermal parameter $\alpha$). The results of our present investigation might be useful for understanding
  different nonlinear electrostatic phenomena in both space (viz. ionosphere and mesosphere) and
  laboratory plasmas (viz. high intensity laser irradiation and hot cathode discharge).
\PACS{
      {PACS-key}{discribing text of that key}   \and
      {PACS-key}{discribing text of that key}
     } % end of PACS codes
} %end of abstract
\authorrunning{M. H. Rahman  \textit{et al.}}
\titlerunning{Modulational instability, rogue waves, and envelope solitons........}
\maketitle
\section{Introduction}
\label{1sec:int}
Recently, the research regarding dusty plasma is one of the fundamental and exponentially growing
branches of plasma physics because of the empirical results directly support the existence of
dust not only in space plasmas (viz. cometary tails, asteroid zones, planetary ring, interstellar medium, lower part
of the earths ionosphere, and the magnetosphere, etc.) but also in laboratory  plasmas (viz. radio frequency
plasma discharge, low temperature physics, and plasma crystals, etc \cite{Mendis1994,Shukla2001,Shukla2002,Verheest2000,Chowdhury2017a,Tasnim2012}.).
Dust acoustic (DA) waves (DAWs) and their associated nonlinear structures (viz. shock, vortices, rogue, and envelope solitons, etc.) are rigorously used by Plasma
physicists to understand the collective behaviours of such kind of dusty plasmas (DP).
Usually, dust grains are assumed negatively charged massive objects due to the collection of electrons from background of plasma species
\cite{Shukla1992,Bharuthram1992}. But a set of policies (e.g. photoemission under ultraviolet
radiation or thermionic emission from grains heated by radiative sources) by which a dust particle
can acquire positive charge, and also lived along with negatively charged dust particles, ions, and electrons in various
DP (viz. upper mesosphere, cometary tails, and Jupiter’s magnetosphere \cite{Mamun2002,Chow1993,Chow1994}, etc.).

Maxwellian distribution, which was developed by Maxwell and Boltzmann, is suitable to describe
the motion of particles in a thermodynamically equilibrium system. But in the astrophysical
objects and space plasmas due to the various physical mechanisms (viz. wave-particle interaction,
particle-particle interaction, and presence of external force field in natural space
plasma environment),  Maxwellian distribution is not adequate to describe the motion of plasma
particles. A number of authors have used non-extensive $q$-distribution \cite{Renyi1955,Tsallis1988,Chowdhury2017b,Tribeche2011}
or non-thermal Cairn's distribution \cite{Cairns1995,Mamun1996,Mamun1997} instead of Maxwellian distribution
to describe the motion of plasma particles in thermodynamically non-equilibrium system. Rogue
waves (RWs), which generate due to the modulational instability (MI) of plane
water waves,  observed first in ocean \cite{Kharif2009}. Nowadays, RWs can be observed in optics,
atmospheric physics, stock market crashes \cite{Yan2010}, super-fluid helium \cite{Ganshin2008},
and in plasma physics \cite{Moslem2011a}, etc. Sayed and Mamun \cite{Sayed2007} examined the presence
of positive dust component significantly modify the fundamental properties of solitary potential structures
along with negative dust in four component DP. El-Taibany \cite{El-Taibany2013} investigated the dependency of DA solitary waves
nature on  opposite charge polarity dust grain masses and temperature. Bains \textit{et al.} \cite{Bains2013} reported
DAWs modulation in the presence of $q$-distributed electrons and ions and found that non-extensive ions have more effects on the MI of the DAWs than electrons.
Misra and Chowdhury \cite{Misra2006} studied MI of DAWs in a DP with non-thermal electrons and ions.
Zaghbeer \textit{et al.} \cite{Zaghbeer2014} examined the effects of $q$-distributed electrons and
ions on DA RWs (DARWs) in opposite polarity DP. Moslem \textit{et al.} \cite{Moslem2011} investigated
the DARWs in a $q$-distributed plasma and found that RWs are influenced by the plasma
parameters. Sultana \textit{et al.} \cite{Sultana2014} studied envelope solitons and their MI in DP
and found that the MI conditions of the modified envelope solitons are also influenced due
to the variation of the intrinsic plasma parameters. Therefore, in our present work, we will examine the
MI of the DAWs propagating in opposite polarity DP
(inertial warm negatively and positively charged dust particles) as well as inertialess $q$-distributed
electrons and non-thermal ions which abundantly occurs in astrophysical environments,
(viz. upper mesosphere, cometary tails, and Jupiter’s magnetosphere, etc).

The present paper is organized as follows: In Sec. \ref{1sec:eqn}, the model equations
and derivation of the NLS equation are presented. In Sec. \ref{1sec:mod}, the MI and
rogue waves are examined. In Sec. \ref{1sec:Env}, the bright and dark envelope solitons
are observed. In sec. \ref{1sec:Dis}, a brief discussion  is provided.
\section{Model equations and derivation of the NLS equation}
\label{1sec:eqn}
\begin{figure}[t!]
\vspace{6cm}
\includegraphics{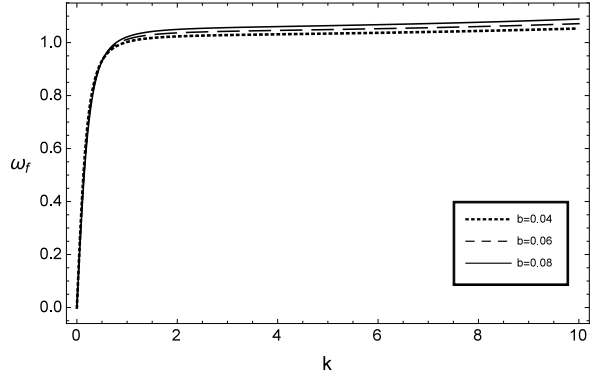}
\caption{The variation of $\omega_f$ with $k$ for different values of $b$; along
 with $a=1.5$, $q=1.3$, $\alpha=0.3$, $\delta=0.4$,  $\mu_i=0.6$,  $\sigma_1=0.0001$, and $\sigma_2=0.001$.}
\label{1Fig:F1}
\end{figure}

\begin{figure}[t!]
\vspace{5cm}
\includegraphics{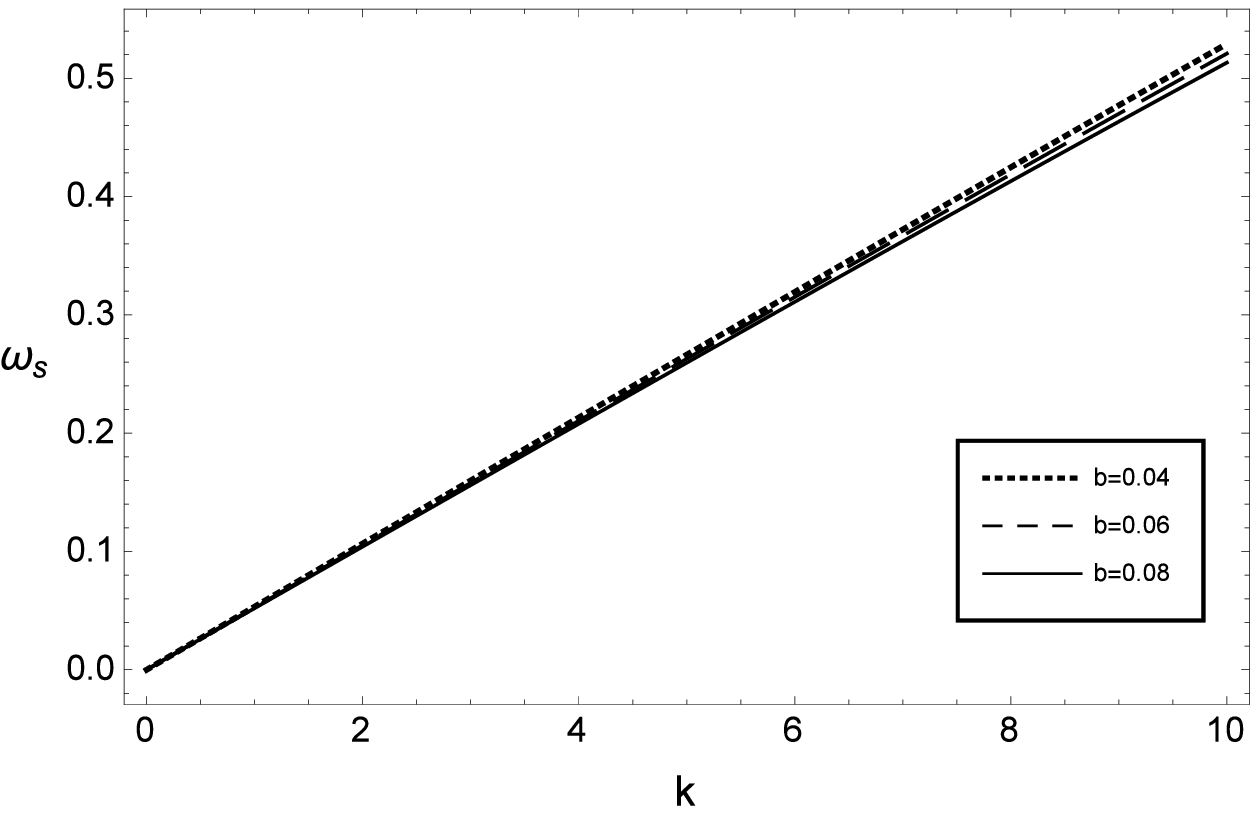}
\caption{The variation of $\omega_s$ with $k$ for different values of $b$;
along with  $a=1.5$, $q=1.3$, $\alpha=0.3$, $\delta=0.4$, $\mu_i=0.6$, $\sigma_1=0.0001$, and $\sigma_2=0.001$.}
\label{1Fig:F2}
\end{figure}

We consider a collisionless, fully ionized, unmagnetized plasma system comprising of
inertial warm negatively  charged dust particles (mass $m_1$; charge $q_1=-Z_1e$), and positively
charged dust particles (mass $m_2$; charge $q_2=+Z_2e$), as well as $q$-distributed electrons
(mass $m_e$; charge $-e$), and non-thermal ions (mass $m_i$; charge $+e$). $Z_1$ ($Z_2$) is the number of electrons (protons)
residing on a negative (positive) dust particle. At equilibrium, the quasi-neutrality
condition can be expressed as $Z_1n_{10} + n_{e0}= Z_2 n_{20}+n_{i0}$;
where $n_{10}$, $n_{e0}$, $n_{20}$, and $n_{i0}$ are the equilibrium number densities of warm negatively
charged dust particles, $q$-distributed electrons, positively charged dust particles, and non-thermal ions,
respectively. The normalized governing equations of the DAWs in our plasma system are
\begin{eqnarray}
&&\hspace*{-1.2cm}\frac{\partial n_1}{\partial t}+\frac{\partial}{\partial x}(n_1 u_1)=0,
\label{1eq:1}\\
&&\hspace*{-1.2cm}\frac{\partial u_1}{\partial t} + u_1\frac{\partial u_1 }{\partial x} + 3\sigma_1  n_1\frac{\partial n_1 }{\partial x}=\frac{\partial \phi}{\partial x},
\label{1eq:2}\\
&&\hspace*{-1.2cm}\frac{\partial n_2}{\partial t}+\frac{\partial}{\partial x}(n_2 u_2)=0,
\label{1eq:3}\\
&&\hspace*{-1.2cm}\frac{\partial u_2}{\partial t} + u_2\frac{\partial u_2 }{\partial x}+ 3\sigma_2  n_2\frac{\partial n_2 }{\partial x}=-a \frac{\partial \phi}{\partial x},
\label{1eq:4}\\
&&\hspace*{-1.2cm}\frac{\partial^2 \phi}{\partial x^2}=(\mu_i+b-1)n_e-\mu_i n_i+n_1-b n_2,
\label{1eq:5}
\end{eqnarray}
where $n_1$ ($n_2$) is the number density of negatively (positively) charged dust particles normalized by its equilibrium value
$n_{10}$ ($n_{20}$); $u_1(u_2)$ is the negatively (positively) charged dust fluid speed
normalized by $C_1=(Z_1 T_i/m_1)^{1/2}$, and the electrostatic wave potential $\phi$
is normalized by $T_i/e$ (with $e$ being the magnitude of an electron charge);
$T_1$, $T_2$, $T_i$, and $T_e$ is the temperature of  negatively charged dust,
positively charged dust, non-thermal ions, and $q$-distributed electrons, respectively;
the time and space variables are normalized by ${\omega^{-1}_{pd1}}=(m_1/4\pi Z^2_1e^2 n_{10})^{1/2}$
and $\lambda_{Dd1}=(T_{i}/4 \pi Z_1 e^2 n_{10})^{1/2}$, respectively; some related parameters are
defined as $a=m_1Z_2/m_2Z_1$, $b=Z_2n_{20}/Z_1 n_{10}$
$\mu_i=n_{i0}/Z_1n_{10}$, $\sigma_1=T_1/Z_1 T_i$, and $\sigma_2=T_2 m_1/Z_1T_i m_2$.
The expression for the number density of non-extensive electrons following the
non-extensive $q$-distribution \cite{Renyi1955,Tsallis1988,Chowdhury2017b}  can be written  as
\begin{eqnarray}
&&\hspace*{-1.8cm}n_e= \left[1+(q-1)\delta\right]^{\frac{q+1}{2(q-1)}}
\label{1eq:6}
\end{eqnarray}
where $\delta=T_i/T_e$ and $q$ is the non-extensive parameter. When $q<1$ ($q>1$) refers to super-extensivity (sub-extensivity) \cite{Tribeche2011} and
$q=1$ refers to Maxwellian \cite{Verheest2000}. The expression for the number density of non-thermal ions
following the non-thermal Cairn's distribution \cite{Cairns1995,Mamun1996,Mamun1997}  can be written  as
\begin{eqnarray}
&&\hspace*{-1.0cm}n_i=(1+\beta\phi+\beta \phi^2)~\mbox{exp}(-\phi),
\label{1eq:7}
\end{eqnarray}
where $\beta=4\alpha/(1+3\alpha)$ with $\alpha$ being the non-thermal parameter. We note that in many space plasma systems contain fraction of energetic or fast
plasma particle in addition to thermal ones. Now, by substituting Eqs. (\ref{1eq:6}) and
(\ref{1eq:7}) into Eq. (\ref{1eq:5}), and expanding up to third order in $\phi$, we get
\begin{eqnarray}
&&\hspace*{0.5cm}\frac{\partial^2 \phi}{\partial x^2}=b-1+n_1-b n_2+\gamma_1 \phi+\gamma_2\phi^2+\gamma_3 \phi^3+\cdot\cdot\cdot\cdot,
\label{1eq:8}
\end{eqnarray}
where
\begin{eqnarray}
&&\hspace*{1.6cm}\gamma_1=\frac{1}{2}[(b+\mu_i-1)(q+1)\delta+2\mu_i(1-\beta)],
\nonumber\\
&&\hspace*{1.6cm}\gamma_2=\frac{1}{8}[(b+\mu_i-1)(q+1)(3-q)\delta^2-4\mu_i],
\nonumber\\
&&\hspace*{1.6cm}\gamma_3=\frac{1}{48}[(b+\mu_i-1)(q+1)(q-3)(3q-5)\delta^3+8\mu_i(1+3\beta)].
\nonumber\
\end{eqnarray}
To study the MI of the DAWs, we will derive the NLS equation by
employing the reductive perturbation method. So, we first introduce
the stretched co-ordinates
\begin{eqnarray}
&&\hspace*{-2.8cm}\xi={\epsilon}(x-v_gt),
\label{1eq:9}\\
&&\hspace*{-2.8cm}\tau={\epsilon}^2 t,
\label{1eq:10}
\end{eqnarray}
where $v_g$ is the envelope group velocity and $\epsilon ~(0<\epsilon<1)$ is a
small (real) parameter. Then, we can write a general expression for the dependent variables as
\begin{eqnarray}
&&\hspace*{-0.2cm}G(x,t)=G_0 +\sum_{m=1}^{\infty}\epsilon^{(m)}\sum_{l=-\infty}^{\infty}G_{l}^{(m)}(\xi,\tau) ~\mbox{exp}(i l\Upsilon),
\label{1eq:11}
\end{eqnarray}
where $G_l^{(m)}=[n_{1l}^{(m)}, u_{1l}^{(m)}, n_{2l}^{(m)}, u_{2l}^{(m)}, \phi_l^{(m)}]^T$,
$G_0=[1, 0, 1, 0, 0]^T$, $\Upsilon=(k x -\omega t)$, and $k$ ($\omega$) is the fundamental carrier wave number (frequency).
All elements of $G_l^{(m)}$ satisfy the reality condition $G_{-l}^{(m)}=G_l^{*(m)}$, where the asterisk indicates the complex conjugate.
The derivative operators in the above equations are treated as follows:
\begin{eqnarray}
&&\hspace*{-1.2cm}\frac{\partial}{\partial t}\rightarrow\frac{\partial}{\partial t}-\epsilon v_g \frac{\partial}{\partial\xi}+\epsilon^2\frac{\partial}{\partial\tau},
\label{1eq:12}\\
&&\hspace*{-1.2cm}\frac{\partial}{\partial x}\rightarrow\frac{\partial}{\partial x}+\epsilon\frac{\partial}{\partial\xi}.
\label{1eq:13}
\end{eqnarray}
Now, by substituting Eqs. (\ref{1eq:9})$-$(\ref{1eq:13}) into Eqs. (\ref{1eq:1})$-$(\ref{1eq:4}), and (\ref{1eq:8}), and equating the coefficients of $\epsilon$ for $m=l=1$, we obtain
\begin{eqnarray}
&&\hspace*{0.2cm}n_{11}^{(1)}=\frac{k^2}{S}\phi_1^{(1)},~~~~~~~ u_{11}^{(1)}=\frac{k\omega}{S}\phi_1^{(1)},
\nonumber\\
&&\hspace*{0.2cm}n_{21}^{(1)}=\frac{a k^2}{A}\phi_1^{(1)},~~~~~u_{21}^{(1)}=\frac{a k\omega}{A}\phi_1^{(1)},
\label{1eq:14}
\end{eqnarray}
where $S=\lambda k^2-\omega^2$, $A=\omega^2-\theta k^2$, $\lambda=3 \sigma_1$, and $\theta=3 \sigma_2$. We thus obtain the dispersion relation for DAWs
\begin{eqnarray}
&&\hspace*{-0.4cm}\omega^2=\frac{k^2M\pm k^2 \sqrt{M^2-4 GH}}{2G},
\label{1eq:15}
\end{eqnarray}
where $M= \theta k^2+\lambda k^2+\theta \gamma_1+\lambda\gamma_1+ab+1,~G= k^2+\gamma_1 $, and
$H= \theta\lambda k^2+\theta\gamma_1\lambda+\theta+ab\lambda.$ To obtain real and positive
values of $\omega$ from Eq. (\ref{1eq:15}), the condition $M^2>4GH$ should be verified. The
positive sign in Eq. (\ref{1eq:15}) corresponds to the fast DA mode ($\omega_f$), whereas
the negative sign corresponds to the slow DA mode ($\omega_s$).
Physically, in fast mode  both dust species oscillate in same phase with electrons and ions.
On the other hand, in slow mode one of the dust species oscillates in opposite phase with electrons, ions, and another dust species.
We have numerically analyzed the $\omega_f$ and $\omega_s$ in Figs. \ref{1Fig:F1} and \ref{1Fig:F2}, which clearly indicate that
(a) as we increase the value of carrier wave number $k$, firstly, the $\omega_f$ increases exponentially
but after a particular value of $k$, $\omega_f$  remains almost constant (please see Fig. \ref{1Fig:F1});
(b) the $\omega_f$ increases with the increase of $Z_2$ for fixed value of $n_{20}$, $Z_1$,  and $n_{10}$ (via $\frac{}{}b=Z_2n_{20}/Z_1n_{10}$);
(c) as we increase the value of $k$, the $\omega_s$ linearly increases (please see Fig. \ref{1Fig:F2});
(d) the $\omega_s$ increases with the increase of $Z_1$ for fixed value of $Z_2$, $n_{20}$,  and $n_{10}$ (via $b=Z_2n_{20}/Z_1n_{10}$). So, the charge of
the dust particles  play an opposite role for the $\omega_f$  and $\omega_s$ of the DAWs, respectively.
The second-order of $\epsilon$ when $(m=2)$ reduced equations with $(l=1)$ are
\begin{eqnarray}
&&\hspace*{-0.005cm}n_{11}^{(2)}=\frac{k^2}{S}\phi_1^{(2)}+\frac{i}{S^2}(\lambda k^3+k\omega^2-2v_g\omega k^2-kS )\frac{\partial \phi_1^{(1)}}{\partial\xi},
\nonumber\\
&&\hspace*{-0.005cm}u_{11}^{(2)}=\frac{k\omega}{S}\phi_1^{(2)}+\frac{i}{S^2}(\lambda\omega k^2+\omega^3-2v_g k\omega^2-v_g k S)\frac{\partial\phi_1^{(1)}}{\partial\xi},
\nonumber\\
&&\hspace*{-0.005cm}n_{21}^{(2)}=\frac{a k^2}{A}\phi_1^{(2)}-\frac{ia}{A^2}(\theta k^3+k\omega^2+k A-2\omega v_g k^2)\frac{\partial\phi_1^{(1)}}{\partial\xi},
\nonumber\\
&&\hspace*{-0.005cm}u_{21}^{(2)}=\frac{a k\omega}{A}\phi_1^{(2)}-\frac{ia}{A^2}(\theta\omega k^2+\omega^3+v_g k A-2v_g k\omega^2)\frac{\partial\phi_1^{(1)}}{\partial\xi},
\nonumber\
\end{eqnarray}
with the compatibility condition
\begin{eqnarray}
&&\hspace*{-3.5cm}v_g=\frac{F1-2 S^2 A^2-SA(A-ab S )}{2k \omega( A^2 + ab S^2 )},
\label{1eq:16}
\end{eqnarray}
where
\begin{eqnarray}
&&\hspace*{-2.3cm}F1=k^2(\lambda A^2 + ab \theta S^2)+ \omega^2 (A^2+ab S^2).
\nonumber\
\end{eqnarray}
The second-order of $\epsilon$ when $(m=2)$ reduced equations with $(l=1)$ are
\begin{eqnarray}
&&\hspace*{-3.3cm}n_{12}^{(2)}=C_1|\phi_1^{(1)}|^2,~~~~n_{10}^{(2)}=C_6 |\phi_1^{(1)}|^2,
\nonumber\\
&&\hspace*{-3.3cm}u_{12}^{(2)}=C_2 |\phi_1^{(1)}|^2,~~~~u_{10}^{(2)}=C_7|\phi_1^{(1)}|^2,
\nonumber\\
&&\hspace*{-3.3cm}n_{22}^{(2)}=C_3|\phi_1^{(1)}|^2,~~~~n_{20}^{(2)}=C_8 |\phi_1^{(1)}|^2,
\nonumber\\
&&\hspace*{-3.3cm}u_{22}^{(2)}=C_4 |\phi_1^{(1)}|^2,~~~~u_{20}^{(2)}=C_9|\phi_1^{(1)}|^2,
\nonumber\\
&&\hspace*{-3.3cm}\phi_2^{(2)}=C_5 |\phi_1^{(1)}|^2~~~~~\phi_0^{(2)}=C_{10} |\phi_1^{(1)}|^2,
\label{1eq:17}
\end{eqnarray}
where
\begin{eqnarray}
&&\hspace*{-1.2cm}C_1=\frac{2C_5 k^2 S^2  -(3 \omega^2 k^4+\lambda k^6)}{2S^3},
\nonumber\\
&&\hspace*{-1.2cm}C_2=\frac{\omega C_1 S^2 -\omega k^4}{k S^2},
\nonumber\\
&&\hspace*{-1.2cm}C_3=\frac{3a^2\omega^2 k^4 +\theta a^2 k^6+2a C_5 A^2 k^2}{2A^3},
\nonumber\\
&&\hspace*{-1.2cm}C_4=\frac{ \omega C_3 A^2-\omega a^2 k^4 }{k A^2},
\nonumber\\
&&\hspace*{-1.2cm}C_5=\frac{F2+b S^3(3a^2 \omega^2 k^4+\theta a^2 k^6)}{2S^2k^2 A^3+2A^3 S^3 (4k^2+\gamma_1)-2ab A^2 k^2S^3},
\nonumber\\
&&\hspace*{-1.2cm}F2=A^3 (3\omega^2k^4+\lambda k^6)-2\gamma_2 A^3 S^3,
\nonumber\\
&&\hspace*{-1.2cm}C_6=\frac{2 v_g \omega k^3+\lambda k^4+k^2\omega^2-C_{10}S^2}{S^2(v^2_g-\lambda)},
\nonumber\\
&&\hspace*{-1.2cm}C_7=\frac{v_g C_6 S^2-2\omega k^3}{S^2},
\nonumber\\
&&\hspace*{-1.2cm}C_8=\frac{2 v_g \omega a^2 k^3+\theta a^2k^4+a^2 k^2 \omega ^2+a C_{10}A^2}{A^2(v^2_g - \theta )},
\nonumber\\
&&\hspace*{-1.2cm}C_9=\frac{v_g C_8 A^2 -2\omega a^2 k^3}{A^2},
\nonumber\\
&&\hspace*{-1.2cm}C_{10} =\frac{2\gamma_2 A^2 S^2(v^2_g -\theta )( v^2_g -\lambda)+F3}{ab A^2 S^2 (v^2_g -\lambda)+F4},
\nonumber\\
&&\hspace*{-1.2cm}F3=A^2( 2 v_g \omega k^3+\lambda k^4+k^2 \omega ^2 )(v^2_g - \theta )
\nonumber\\
&&\hspace*{-0.5cm}-b S^2( 2 v_g \omega a^2 k^3+\theta a^2 k^4+a^2 k^2 \omega ^2)(v_g^2 -\lambda),
\nonumber\\
&&\hspace*{-1.2cm}F4=A^2 S^2 (v^2_g - \theta)- \gamma_1 A^2 S^2(v^2_g - \theta)(v^2_g -\lambda).
\nonumber\
\end{eqnarray}
Finally, the third harmonic modes $(m=3)$ and $(l=1)$ and  with the help of  Eqs. $(\ref{1eq:14})$-$(\ref{1eq:17})$,
give a system of equations, which can be reduced to the following  NLS equation
\begin{eqnarray}
&&\hspace*{-3.0cm}i\frac{\partial \Phi}{\partial \tau}+P\frac{\partial^2 \Phi}{\partial \xi^2}+Q|\Phi|^2\Phi=0,
\label{1eq:18}
\end{eqnarray}
where $\Phi=\phi_1^{(1)}$ for simplicity. The dispersion coefficient P
is given by
\begin{eqnarray}
&&\hspace*{-3.0cm}P =\frac{F5-A^3S^3}{2 AS\omega k^2(A^2+ab S^2)},
\nonumber\
\end{eqnarray}
where
\begin{eqnarray}
&&\hspace*{0.7cm}F5=(v_g \omega A^3 -\lambda k A^3)(\lambda k^3-2\omega v_g k^2+k \omega^2-kS)
\nonumber\\
&&\hspace*{1.5cm}+(v_g k A^3-\omega A^3)(\lambda \omega k^2 - 2v_g k \omega^2+\omega^3-k v_g S)
\nonumber\\
&&\hspace*{1.5cm}-ab S^3(v_g \omega -\theta k)(\theta k^3-2\omega v_g k^2+k\omega^2+kA)
\nonumber\\
&&\hspace*{1.5cm}-ab S^3(v_g k-\omega)(\theta\omega k^2 - 2v_g k\omega^2+\omega^3+kv_gA).
\nonumber\
\end{eqnarray}
The nonlinear coefficient Q is given by
\begin{eqnarray}
&&\hspace*{-1.2cm}Q=\frac{A^2S^2\{2\gamma_2(C_5+C_{10})+3\gamma_3\}-F6}{2\omega k^2(A^2+ab S^2)},
\nonumber\
\end{eqnarray}
where
\begin{eqnarray}
&&\hspace*{2.0cm}F6=2\omega A^2 k^3(C_2+C_{7})+A^2(\omega^2k^2+\lambda k^4)(C_1+C_6)
\nonumber\\
&&\hspace*{2.6cm}+2ab \omega S^2 k^3(C_4+C_9)+S^2(ab  k^2\omega^2+ab \theta k^4)(C_3+C_8).
\nonumber\
\end{eqnarray}
\begin{figure}[t!]
\vspace{6cm}
\includegraphics{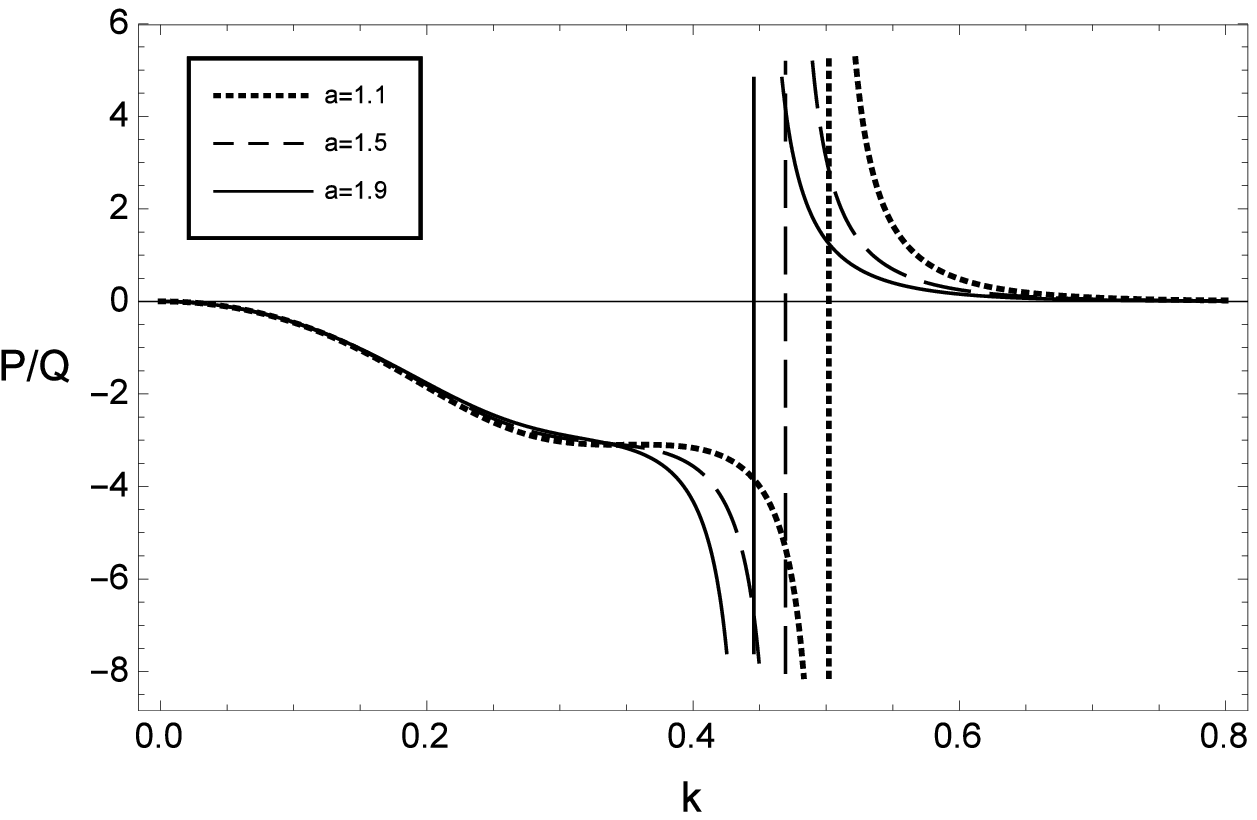}
\caption{The variation of  $P/Q$ with $k$ for different values of $a$; along with
$b=0.08$,  $q=1.3$, $\alpha=0.3$, $\delta=0.4$, $\mu_i=0.6$,  $\sigma_1=0.0001$, $\sigma_2=0.001$, and $\omega_f$.}
\label{1Fig:F3}
\end{figure}
\begin{figure}[t!]
\vspace{6cm}
\includegraphics{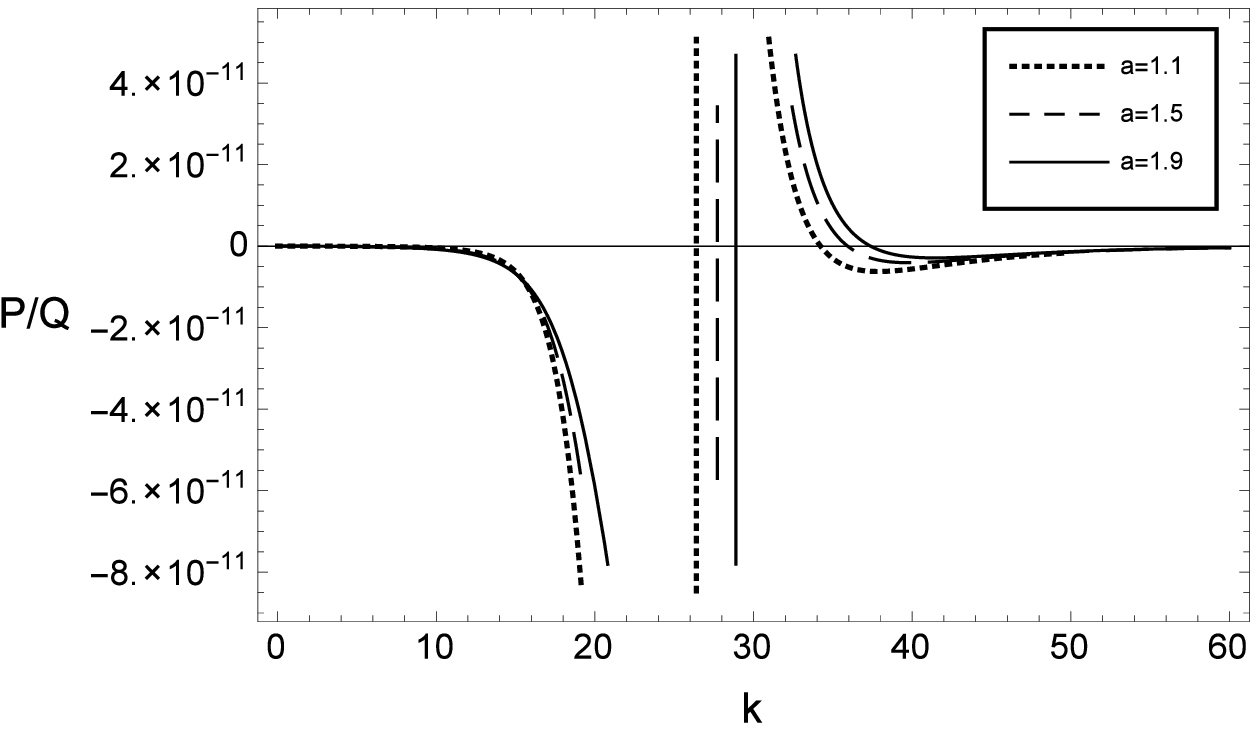}
\caption{The variation of  $P/Q$ with $k$ for different values of $a$; along with
$b=0.08$, $q=1.3$, $\alpha=0.3$, $\delta=0.4$, $\mu_i=0.6$, $\sigma_1=0.0001$, $\sigma_2=0.001$, and $\omega_s$.}
\label{1Fig:F4}
\end{figure}
\section{Modulational instability and rogue waves}
\label{1sec:mod}
\begin{figure}[t!]
\vspace{6cm}
\includegraphics{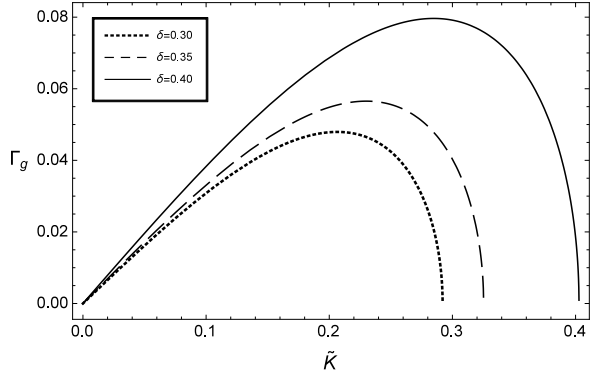}
\caption{Plot of the $\Gamma_g$ with $\tilde{k}$ for $\delta$; along with $k=0.5$, $\phi_0=0.5$,
$a=1.5$, $b=0.08$, $q=1.3$, $\alpha=0.3$, $\mu_i=0.6$, $\sigma_1=0.0001$, $\sigma_2=0.001$, and $\omega_f$.}
\label{1Fig:F5}
\end{figure}

\begin{figure}[t!]
\vspace{6cm}
\includegraphics{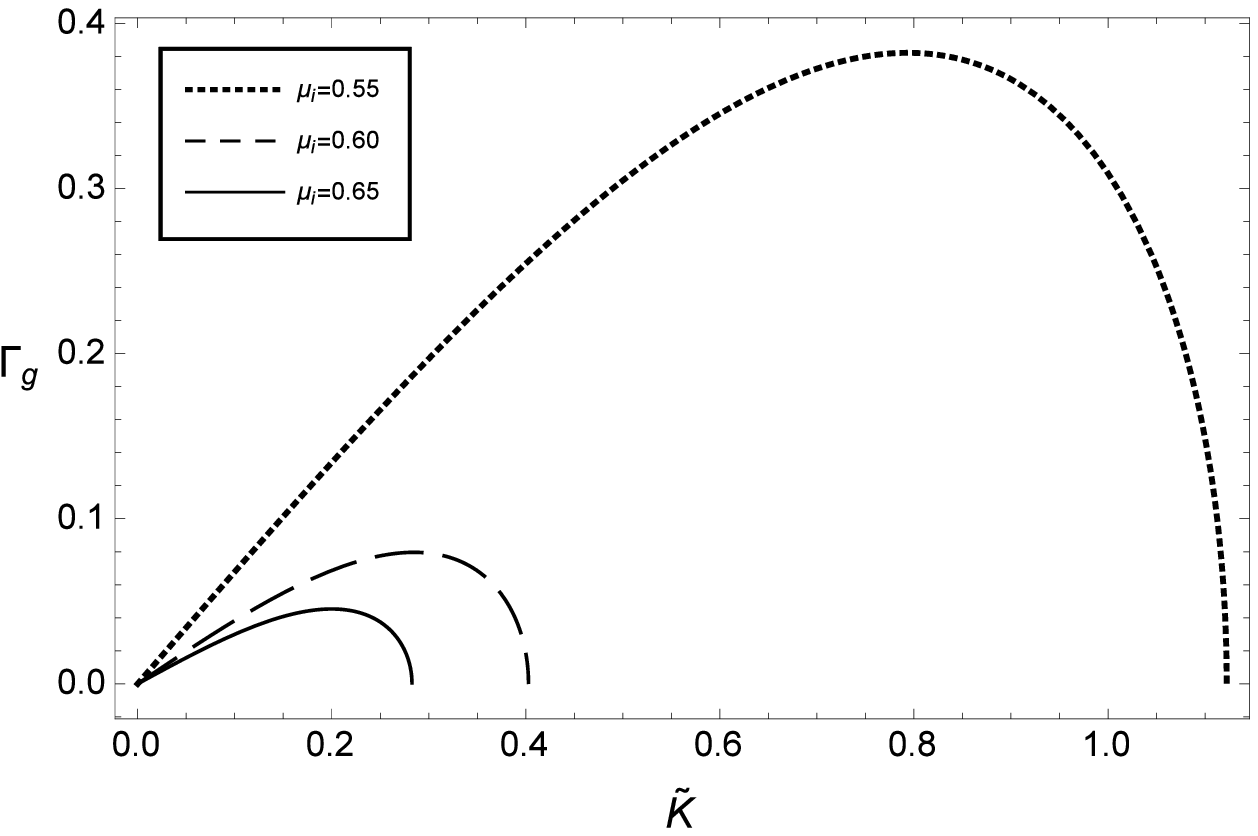}
\caption{Plot of the $\Gamma_g$ with $\tilde{k}$ for $\mu_i$; along with $k=0.5$, $\phi_0=0.5$,
$a=1.5$, $b=0.08$, $q=1.3$, $\alpha=0.3$, $\delta=0.4$,  $\sigma_1=0.0001$,  $\sigma_2=0.001$, and $\omega_f$.}
\label{1Fig:F6}
\end{figure}

\begin{figure}[t!]
\vspace{6cm}
\includegraphics{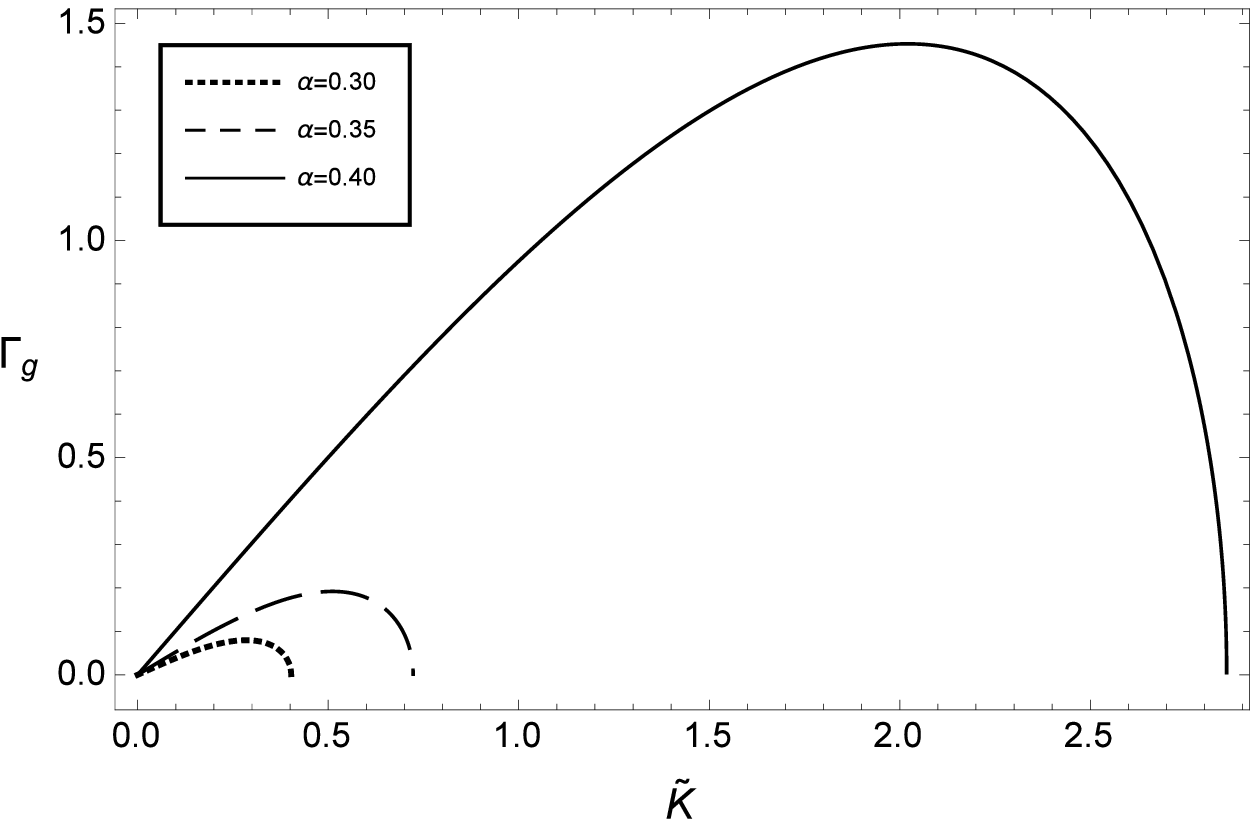}
\caption{Plot of the $\Gamma_g$ with $\tilde{k}$ for $\alpha$; along with $k=0.5$, $\phi_0=0.5$,
$a=1.5$, $b=0.08$, $q=1.3$, $\delta=0.4$, $\mu_i=0.6$, $\sigma_1=0.0001$, $\sigma_2=0.001$, and $\omega_f$.}
\label{1Fig:F7}
\end{figure}

To study the MI of DAWs, we consider the linear solution of the NLS equation (\ref{1eq:18})
in the form $\Phi=\hat{\Phi}e^{iQ|\hat{\Phi}|^2 \tau}+c.~c$ (c. c denotes the
complex conjugate), where $\hat{\Phi}=\hat{\Phi}_0+\epsilon \hat{\Phi}_1$ and
$\hat{\Phi}_1=\hat{\Phi}_{1,0}e^{i(\tilde{k}\xi-\tilde{\omega}\tau)}$+c. c
(the perturbed wave number $\tilde{k}$ and the frequency $\tilde{\omega}$ are
different from $k$ and $\omega$). Now, by substituting these values in
Eq. (\ref{1eq:18}) the following nonlinear dispersion relation can be obtained as
\cite{Chowdhury2018,Sultana2011,Kourakis2005,Fedele2002a,Fedele2002b}
\begin{eqnarray}
&&\hspace*{-1.2cm}\tilde{\omega}^2=P^2\tilde{k}^2 \left(\tilde{k}^2-\frac{2|\hat{\Phi}_0|^2}{P/Q}\right).
\label{1eq:19}
\end{eqnarray}
When $P/Q>0$, the DAWs are modulationally unstable  against external perturbation, and in this region rogue waves and bright envelope
solitons exist. On the other hand, when $P/Q<0$, the DAWs are modulationally stable and in this region
dark envelope solitons exist. When $P/Q\rightarrow\pm\infty$, the corresponding value of $k$ ($=k_c$)
is called critical or threshold wave number for the onset of MI. We have graphically shown how the
ratio of P/Q varies with $k$ for different values of $a$ in Fig. \ref{1Fig:F3} and \ref{1Fig:F4}, respectively.
It is obvious from Fig. \ref{1Fig:F3}  and \ref{1Fig:F4}  that (a) for both $\omega_f$ and $\omega_s$, there is a stable/unstable region occurred for DAWs (see Fig. \ref{1Fig:F3} and \ref{1Fig:F4}); (b) DAWs are modulationally stable (unstable) for long (short) wavelength; (c) the $k_c$ value decreases (increases), as we
increase the value of $m_1$ ($m_2$) for fixed values of $Z_2$ and $Z_1$ (via $a=m_1Z_2/m_2Z_1$); (d) on the other hand, the $k_c$  increases (decreases), as we
increase the value of $m_1$ ($m_2$) for fixed values of $Z_2$ and $Z_1$ (via $a=m_1Z_2/m_2Z_1$). When simultaneously $P/Q>0$
and $\tilde{k}<\tilde{k}_c=(2 P|\hat{\Phi}_0|/Q)^{1/2}$, from Eq. (\ref{1eq:19}) the growth
rate ($\Gamma_g$) of  MI can be written as
\begin{eqnarray}
&&\hspace*{-1.2cm}\Gamma_g=|P|\tilde{k}^2\left(\frac{\tilde{k}^2_c}{\tilde{k}^2}-1\right)^{1/2}.
\label{1eq:20}
\end{eqnarray}
We have graphically shown how the $\Gamma_g$ varies with $\tilde{k}$ for different values of $\delta$, $\mu_i$,
and $\beta$ in  Figs. \ref{1Fig:F5}$-$\ref{1Fig:F7}. It is obvious from  Figs. \ref{1Fig:F5}$-$\ref{1Fig:F7}
that (a) the  $\Gamma_g$ increases (decreases) with the increase of $T_i$ ($T_e$); (b) on the other hand, the $\Gamma_g$
increases with the decrease of $n_{i0}$  for fixed value of $Z_1$ and $n_{10}$; (c) as we increases the
value of $\beta$, the $\Gamma_g$ increases (see Fig. \ref{1Fig:F7}). The physics of this result is that, the  maximum value of the
growth rate of DAWs increases since the nonlinearity increases with the increase of the value
of $\beta$. So, more non-thermal ions are used to enhance the maximum value of the growth rate.
In the unstable region ($P/Q>0$), the NLS equation (\ref{1eq:18}) has rogue wave (rational)
solution, which can be written as \cite{Akhmediev2009,Ankiewiez2009}
\begin{eqnarray}
&&\hspace*{-1.2cm}\Phi(\xi,\tau)=\sqrt{\frac{2P}{Q}} \left[\frac{4(1+4iP\tau)}{1+16P^2\tau^2+4\xi^2}-1 \right]\mbox{exp}(i2P\tau).
\label{1eq:21}
\end{eqnarray}
\begin{figure}[t!]
\vspace{6cm}
\includegraphics{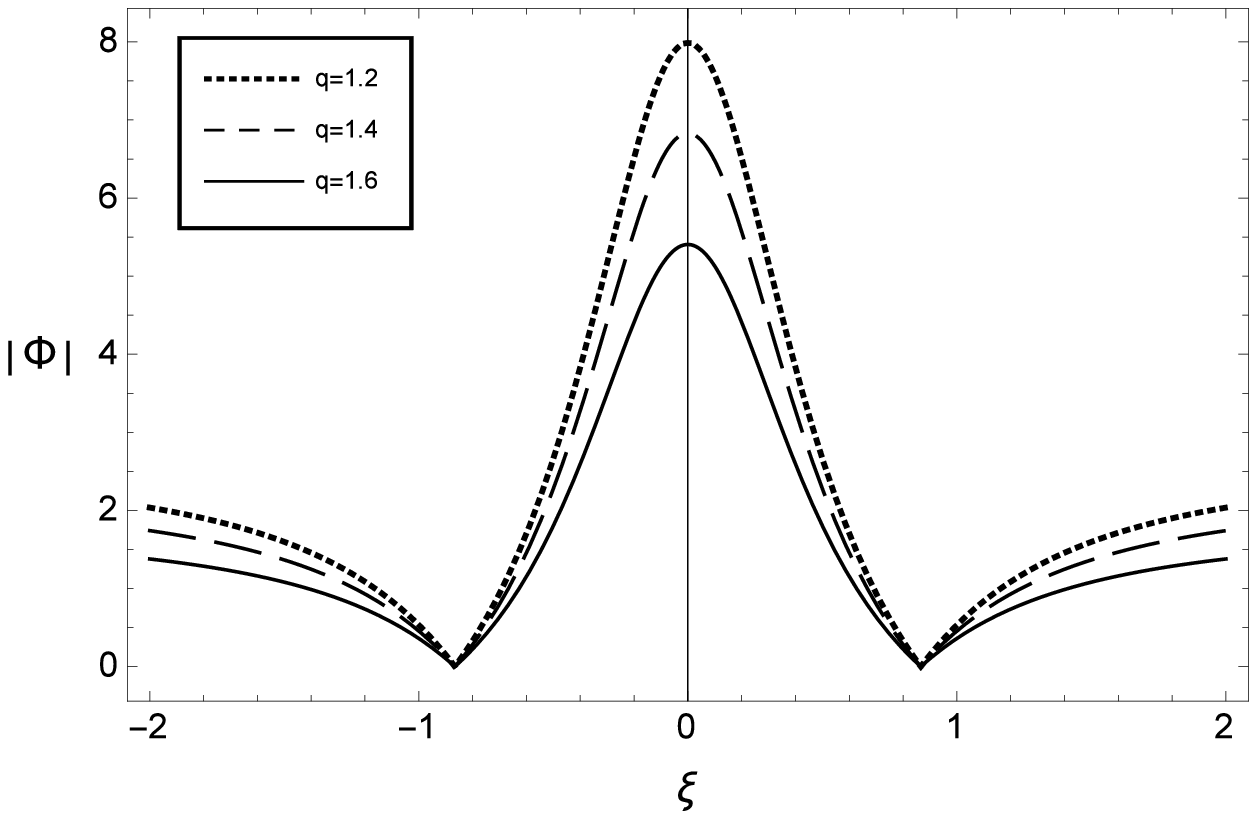}
\caption{The variation of $|\Phi|$ with $\xi$ for q = positive; along with $k=0.5$, $\tau=0$, $a=1.5$, $b=0.08$,
$\alpha=0.3$, $\delta=0.4$, $\mu_i=0.6$, $\sigma_1=0.0001$, $\sigma_2=0.001$, and $\omega_f$.}
\label{1Fig:F8}
\end{figure}
\begin{figure}[t!]
\vspace{6cm}
\includegraphics{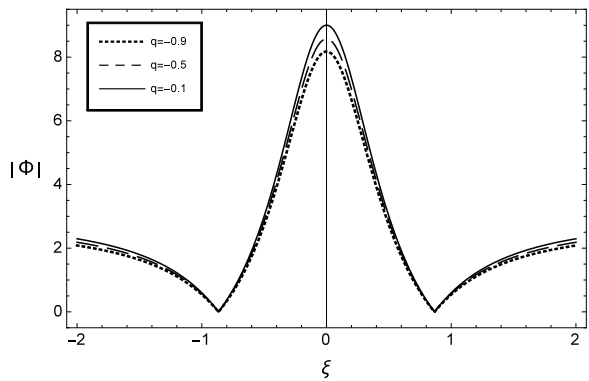}
\caption{The variation of $|\Phi|$ with $\xi$ for q = negative; along with  $k=0.5$, $\tau=0$,  $a=1.5$, $b=0.08$,
$\alpha=0.3$, $\delta=0.4$, $\mu_i=0.6$, $\sigma_1=0.0001$, $\sigma_2=0.001$, and $\omega_f$.}
\label{1Fig:F9}
\end{figure}
The solution (\ref{1eq:21}) predicts the  concentration of large amount of energy of the DAWs  into a small, tiny
region (please see Figs. \ref{1Fig:F8} and \ref{1Fig:F9}) that is caused by the nonlinear behavior of the plasma medium.
It  can be seen from Fig. \ref{1Fig:F8} and \ref{1Fig:F9} that (a) as we increase the value of $q$ in sub-extensive (super-extensive)
limit of the $q$, the amplitude and width of the rogue waves decrease (increase). Generally, when the nonlinearity of the plasma system increases, then excess
nonlinearity leads to generate more energetic, taller  rogue waves, by concentrating a reasonable amount of energy into tiny region.
\section{Envelope solitons}
\label{1sec:Env}
There are two types of envelope solitonic solutions exist, namely, bright
and dark envelope solitons, depending on the sign of the coefficients $P$ and $Q$.
\subsection{Bright envelope solitons}
\begin{figure}[t!]
\vspace{6cm}
\includegraphics{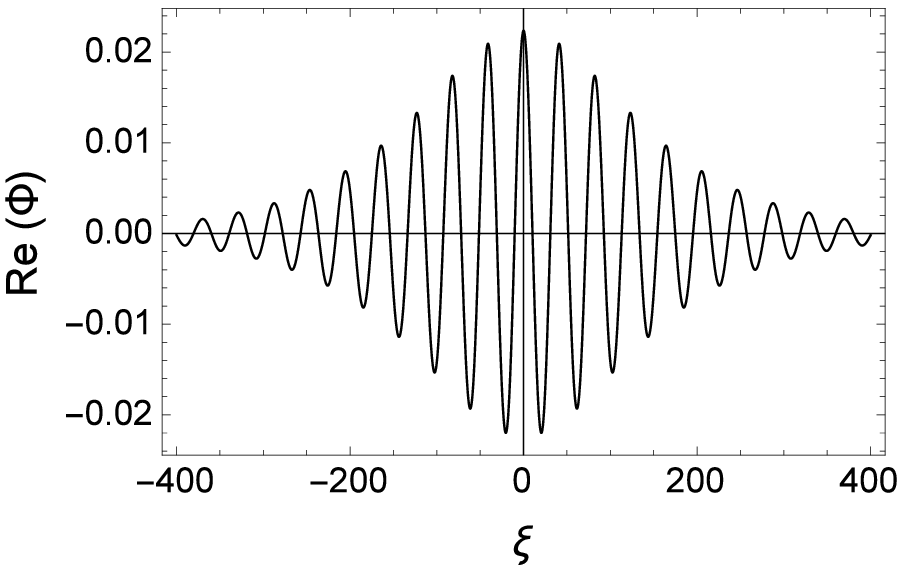}
\caption{The variation of $Re(\Phi)$ with $\xi$ for bright envelope solitons;
along with $k=0.5$, $\tau=0$, $\psi_{0}=0.0005$, $U=0.3$, $\Omega_0=0.4$, $a=1.5$, $b=0.08$, $q=1.3$, $\alpha=0.3$, $\delta=0.4$, $\mu_i=0.6$, $\sigma_1=0.0001$, $\sigma_2=0.001$, and $\omega_f$.}
\label{1Fig:F10}
\end{figure}

When $P/Q>0$, the expression of the bright envelope solitonic solution of Eq. (\ref{1eq:18}) can
be written in the given form \cite{Chowdhury2018,Sultana2011,Kourakis2005,Fedele2002a,Fedele2002b}
\begin{eqnarray}
&&\hspace*{-1.5cm}\Phi(\xi,\tau)=\left[\psi_0~\mbox{sech}^2 \left(\frac{\xi-U\tau}{W}\right)\right]^{1/2}\times \exp \left[\frac{i}{2P}\left\{U\xi+\left(\Omega_0-\frac{U^2}{2}\right)\tau \right\}\right].
\label{1eq:22}
\end{eqnarray}
where $\psi_0$ indicates the envelope amplitude, $U$ is the travelling speed of
the localized pulse, $W$ is the pulse width, which can be written as $W=(2P\psi_0/Q)^{1/2}$, and $\Omega_0$ is the
oscillating frequency for $U=0$. The bright envelope soliton [obtained from
Eq. (\ref{1eq:22})] is depicted in Fig. \ref{1Fig:F10}. It may be noted here the width
of the bright envelope solitons decreases (increases) with the increase of $T_i$ ($T_e$)
but their amplitude remains constant (via $\delta=T_i/T_e$).
\subsection{Dark envelope solitons}
\begin{figure}[t!]
\vspace{5cm}
\includegraphics{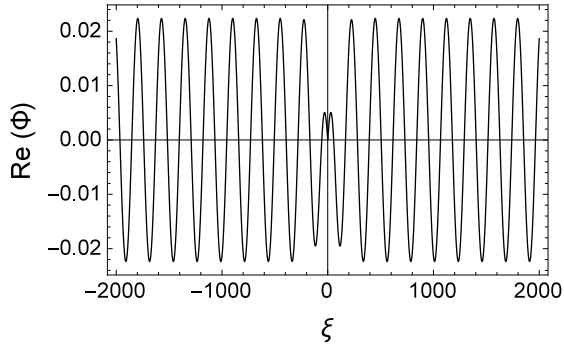}
\caption{The variation of $Re(\Phi)$ with $\xi$ for dark envelope solitons;
along with $k=0.2$, $\tau=0$, $\psi_{0}=0.0005$, $U=0.3$, $\Omega_0=0.4$, $a=1.5$, $b=0.08$, $q=1.3$, $\alpha=0.3$,
$\delta=0.4$, $\mu_i=0.6$, $\sigma_1=0.0001$, $\sigma_2=0.001$, and $\omega_f$.}
\label{1Fig:F11}
\end{figure}

As we know before that the condition for dark envelope soliton is $P/Q<0$.
So, the dark envelope soliton solution of Eq. (\ref{1eq:18}) can be  written as
\cite{Chowdhury2018,Sultana2011,Kourakis2005,Fedele2002a,Fedele2002b}
\begin{eqnarray}
&&\hspace*{-1.0cm}\Phi(\xi,\tau)=\left[\psi_0~\mbox{tanh}^2 \left(\frac{\xi-U\tau}{W}\right)\right]^{1/2}\times \exp \left[\frac{i}{2P}\left\{U\xi-\left(\frac{U^2}{2}-2 P Q \psi_0\right)\tau \right\}\right].
\label{1eq:23}
\end{eqnarray}
The dark envelope soliton [obtained from Eq. (\ref{1eq:23})] is depicted in Fig. \ref{1Fig:F11}.
It may be noted here the width of the dark envelope solitons increases (decreases)
with the increase of $T_i$ ($T_e$) but their amplitude remains constant (via $\delta=T_i/T_e$).
\section{Discussion}
\label{1sec:Dis}
The amplitude modulation of DAWs structures has been theoretically investigated in an
unmagnetized four component opposite polarity DP consisting of inertial warm positively
and negatively charged dust particles as well as non-extensive electrons and non-thermal ions.
A NLS equation, which governs the MI of DAWs and formation of associated rogue waves and
bright envelope solitons in  the unstable regimes, is derived by using reductive perturbation method.
The results that have been found from our investigation can be summarized as follows
\begin{enumerate}
\item{In the fast mode, both dust species oscillate in same phase with electrons and ions.
  On the other hand, in the slow mode one of the dust species oscillate in opposite phase with electrons, ions, and also another dust species.}
\item{For both $\omega_f$ and $\omega_s$, there is a stable/unstable region occurred for DAWs. DAWs are modulationally stable (unstable) for long (short) wavelength.}
\item{The maximum value of the $\Gamma_g$ increases with the decrease of $n_{i0}$  for fixed value of $Z_1$ and $n_{10}$ (via $\mu_i=n_{i0}/Z_1n_{10}$). The growth rate  increases with the $\beta$.}
\item{As we increase the value of $q$ in sub-extensive (super-extensive) limit of the $q$, the amplitude and width of the rogue waves decrease (increase).}
\end{enumerate}
The findings of our present investigation, which is useful to understand the nonlinear
phenomena (viz. rogue waves and envelope solitons) in space DP (viz. ionosphere and mesosphere \cite{Shukla2002}) and laboratories plasmas (viz. high intensity laser
irradiation and hot cathode discharge, etc. \cite{Shukla2002}) where $q$-distributed electrons and non-thermal ions as well
as opposite polarity charged massive dust components are simultaneously co-exist.
\section{Acknowledgement}
M. H. Rahman is grateful to the Bangladesh Ministry of Science and Technology for
awarding the National Science and Technology (NST) Fellowship.

\end{document}